# An optimal method of moments to measure the charge asymmetry at the $Z^0$

N.C. Brümmer
NIKHEF-H
P.O. Box 41882
1009 DB Amsterdam NL

**Abstract**

Parity violation at LEP or SLC can be measured through the charge asymmetry. An optimal method of moments is developed here to measure this asymmetry, as well as similar asymmetries. This method is equivalent to the likelihood fit. It is simpler in use, as it gives analytical formulas for both the asymmetry and its statistical error. These formulas give the dependence of the accuracy on the experimenatal angular acceptance explicitly.

# 1 Introduction

Parity violation occurs both in the production and decay of the $Z^0$ boson at an $e^+e^-$ collider. This results in an asymmetry in the angular distribution of the produced fermions: the forward-backward asymmetry, $A_{FB}$. Equivalently, one can measure an asymmetry between particles and anti-particles: the charge asymmetry, $A_C$.

The conventional way to define this asymmetry is by counting particles and anti-particles in the forward and backward hemispheres as defined with respect to the direction of the incoming electron:

$$A_{FB} = \frac{N^+_{\text{forward}} - N^+_{\text{backward}}}{N^+_{\text{forward}} + N^+_{\text{backward}}} = \frac{N^+_{\text{forward}} - N^-_{\text{forward}}}{N^+_{\text{forward}} + N^-_{\text{forward}}} = A_C \qquad (1)$$

The equality of the two asymmetries can be understood as a consequence of $CP$ invariance.

Although the definition of these asymmetries is clear, a direct application of this definition is not the best way to measure them. The reason is that the data in different angular regions should be weighted differently, according to the changing sensitivity. To see this, we have to look at the explicit angular distribution.

A 'charge', $C = \pm 1$, can be assigned to the fermion and the anti-fermion of the pair created in the $Z^0$ decay. The angular distribution can be given as follows:

$$N^C \propto \frac{3}{8}(1 + \cos^2\theta) + AC\cos\theta \qquad (2)$$

where the polar angle $\theta$ is defined with respect to the incoming electron. This distribution depends only on the combination $C\cos\theta$. It is easy to check that the asymmetry parameter $A$ is equal to the definitions given above for $A_{FB}$ and $A_C$.

From this distribution it can be seen that the sensitivity to $A$ increases with increasing values of $|\cos\theta|$. The standard way to optimise measurements is to use an unbinned log(likelihood) fit [1]. This is a method that will give the best possible accuracy. In this article we will derive a simpler method that is based on the moments of a special variable. This method of moments is equivalent to a likelihood fit, but it gives explicit analytical formulas for both the asymmetry and its statistical error. The dependence on the angular acceptance of the experiment can thus be given analytically.

We will first derive the method of moments for a more general case. Subsequently, the interesting formulas for our explicit problem of $A_{FB} = A_C$ will be given. By construction, the result of the method will not change due to the experimental detection efficiency, even if the efficiency varies (slowly) as a function of the angle $\theta$. It is also possible to treat the effects of a background. Finally a Monte Carlo program was used to compare this method to an unbinned likelihood fit.

# 2 The method of moments

Expressing the asymmetry and its statistical accuracy directly in terms of moments has the advantage of avoiding the requirement to bin the data. However, to derive the best suited moment we will start from the fact that the charge asymmetry can be measured in any bin of $\theta$ separately. Subsequently, we will combine the measurement of many bins in $\theta$ with the proper weights, and express the result in terms of moments. To do this we will switch from taking the average over bins to taking the average over events.

The mathematical derivation in this section will actually be done for a more general angular distribution. Distribution (2) it is a function of the combination $C\cos\theta$. The most general distribution with this property can be described by ($C = \pm 1$):

$$N^C(\cos\theta) \propto f(\cos\theta) + ACg(\cos\theta) \tag{3}$$

where the functions $f$ and $g$ have to satisfy the following symmetry properties:

$$f(-x) = f(x) \qquad g(-x) = -g(x) \tag{4}$$

such that $g(C\cos\theta) = Cg(\cos\theta)$ and $f(C\cos\theta) = f(\cos\theta)$. The asymmetry parameter $A$ is normalised by requiring $\int_{-1}^{+1} |g(x)|dx = \int_{-1}^{+1} f(x)dx$.

The moments of the distribution of a given observable are defined as the averages of powers of this observable. In the case at hand an observable can be any function of the charge $C$ and the angle $\theta$, taking on the value $O_i = O(\theta_i, C_i)$ for each event $i$. Let $O_\theta$ be the average of $O$ inside an angular bin centred at $\theta$. Now the average $\langle O \rangle$ can be defined by a sum over events, but also by a sum over the averages $O_\theta$:

$$\langle O \rangle = \sum_{\text{all } i} O(\theta_i)/N_{\text{tot}} = \sum_\theta O_\theta \cdot N_\theta/N_{\text{tot}} \tag{5}$$

The subscript $\theta$ denotes that the sums are over events $i$ in the bin centred at the angle $\theta$. $N_\theta$ is the number of events in the bin at $\theta$, while $N_{\text{tot}}$ is the total number of events. The $n$-th algebraic moment of $O$ is defined to be $\langle O^n \rangle$, while the corresponding central moment is $\langle [O - \langle O \rangle]^n \rangle$.

To derive the optimal moments to use, we will start by examining the data in different bins of $\theta$. In every bin of $\theta$, the following quantity can be measured:

$$a_\theta = \frac{N_\theta^+ - N_\theta^-}{N_\theta^+ + N_\theta^-} = \frac{\sum_{i \text{ at } \theta} C_i}{\sum_{i \text{ at } \theta} 1} \tag{6}$$

In such a ratio the detection efficiency cancels for each bin separately.

Equation (3) relates a measurement $a_\theta$ in a bin centred at $\theta$ directly to the asymmetry parameter $A$:

$$a_\theta = Ah(\theta) \quad , \text{ with } \quad h(\theta) \equiv g(\theta)/f(\theta) \tag{7}$$

To derive an expression for $A$ in terms of moments we will determine the weighted average of the measurements of $A$ in all the bins. When the theoretical distribution is used to obtain this value, the result is an equation defining $A$ implicitly. By making some simple and plausible assumptions, one can subsequently express the weighted average and its statistical accuracy in terms of moments and get rid of the binning.

Let $A_\theta = a_\theta/h(\theta)$ denote the measurement of $A$ in a given bin. The weighted mean $\overline{A}$ of the values $A_\theta$ and its error are defined by:

$$\overline{A} = \frac{\sum_\theta w_\theta A_\theta}{\sum_\theta w_\theta} \pm \left(\sum_\theta w_\theta\right)^{-1/2} \qquad w_\theta = (\delta A_\theta)^{-2} \tag{8}$$

where the weights $w_\theta$ depend on $\delta A_\theta$, the standard deviation of $A_\theta$:

$$(\delta A_\theta)^2 = \frac{(\delta a_\theta)^2}{h(\theta)^{-2}} = \frac{4N_\theta^+ N_\theta^-}{N_\theta^3 h^2} = \frac{1 - a_\theta^2}{N_\theta h^2} = \frac{1 - A_\theta^2 h^2}{N_\theta h^2} \tag{9}$$

To simplify the denominators of the weights in expression (8) we substitute the measurement $A_\theta$ by the constant $A$. Then definitions (6) can be used to rewrite sums over $\theta$ in terms of sums over events $i$. Subsequently, definition (5) shows how to express the resulting sums as averages.

$$\sum_\theta w_\theta A_\theta = \sum_\theta \frac{N_\theta a_\theta h}{1 - A^2 h^2} = \sum_{\text{all } i} \frac{C_i h(\theta_i)}{1 - A^2 h(\theta_i)^2} = N_{\text{tot}} \left\langle \frac{Ch}{1 - A^2 h^2} \right\rangle \tag{10}$$

$$\sum_\theta w_\theta = \sum_\theta \frac{N_\theta h^2}{1 - A^2 h^2} = \sum_{\text{all } i} \frac{h(\theta_i)^2}{1 - A^2 h(\theta_i)^2} = N_{\text{tot}} \left\langle \frac{h^2}{1 - A^2 h^2} \right\rangle \tag{11}$$

In the weighted mean the factors $N_{\text{tot}}$ cancel to give:

$$\overline{A} = \left\langle \frac{Ch}{1 - A^2 h^2} \right\rangle \cdot \left\langle \frac{h^2}{1 - A^2 h^2} \right\rangle^{-1} \pm N_{\text{tot}}^{-1/2} \left\langle \frac{h^2}{1 - A^2 h^2} \right\rangle^{-1/2} \tag{12}$$

For sufficiently small $|A|$ we can expand the right-hand side of eq. (12) in orders of $A^2$, dropping all terms $\mathcal{O}(A^4)$:

$$\overline{A} = \frac{\langle Ch \rangle}{\langle h^2 \rangle} + A^2 \frac{\langle h^2 \rangle \langle Ch^3 \rangle - \langle Ch \rangle \langle h^4 \rangle}{\langle h^2 \rangle^2} \pm N_{\text{tot}}^{-1/2} \left( \langle h^2 \rangle + A^2 \langle h^4 \rangle \right)^{-1/2} + \mathcal{O}(A^4) \tag{13}$$

This equation can be solved by substituting the first order solution $A = \langle Ch \rangle / \langle h^2 \rangle$ on the right-hand side.

It can be shown that the second term on the right-hand side of equation (13) vanishes. This follows from identities for the distribution (3) that can be derived after invoking the symmetry properties (4) of $f$ and $g$: For positive integers $k$ and $n$, we have:

$$\int_{-1}^{1} f^k(\cos\theta) g^{2n+1}(\cos\theta) \, d\cos\theta = 0 \quad \Rightarrow \quad \langle Ch^{2k+1} \rangle = A \langle h^{2k+2} \rangle \tag{14}$$

Substituting the result for $A$ in the error estimate, equation (13) is now simplified to:

$$\overline{A} = \frac{\langle Ch \rangle}{\langle h^2 \rangle} \pm N_{\text{tot}}^{-1/2} \left( \langle h^2 \rangle + \langle Ch \rangle^2 \langle h^4 \rangle / \langle h^2 \rangle^2 \right)^{-1/2} + \mathcal{O}(A^4) \tag{15}$$

This formula already includes effects of a reduced acceptance, or even the effects of a detection efficiency that is non-uniform in $\cos\theta$, but symmetric in the charge $C$.

It seems as if nothing is left of the binning we started with. However, the above derivation did rely on the existence of sufficient statistics in each bin of $\cos\theta$ for efficiency effects to cancel in the measured ratio $a_\theta = (N_\theta^+ - N_\theta^-)/(N_\theta^+ + N_\theta^-)$. It is thus advisable to reject values of $\theta$ with a very low efficiency.

## 3 The accuracy and the angular acceptance at the $Z^0$

Equation (15) provides a simple formula for the expected accuracy of a measurement of $A_{FB}$ at the $Z^0$ with a reduced angular acceptance, given by $|\cos(\theta)| < \cos\theta_{\min}$. Since we know that $A \lesssim 0.1$, we can now also drop the $\mathcal{O}(A^2)$ term in the error estimate. The function $h = g/h$ for the distribution (2) is equal to:

$$h(\cos\theta) = \frac{8}{3} \cdot \frac{\cos\theta}{(1+\cos^2\theta)} \tag{16}$$

After normalising distribution (2) inside the acceptance, the standard deviation of $A$ is found to be:

$$\sigma(A)^2 \approx \frac{1}{\langle h^2\rangle \cdot N_{\text{tot}}} = \frac{3}{64} \cdot \frac{\cos\theta_{\min}(3+\cos^2\theta_{\min})}{\cos\theta_{\min} - \arctan\cos\theta_{\min}} \cdot \frac{1}{N_{\text{tot}}} \tag{17}$$

Here $N_{\text{tot}}$ stands for the number of events inside the angular acceptance. Equation (17) is visualised in figure 1(a) and can be used to find the expected statistical error for a given angular acceptance and a given number of events inside the acceptance.

To understand the importance of a large polar acceptance when measuring $A_{FB}$, one should also express the accuracy in terms of a the number of $Z^0$ decays $N_Z$. In this case distribution (2) should be normalised for the full acceptance:

$$\sigma(A)^2 \propto \frac{1}{\cos\theta_{\min} - \arctan\cos\theta_{\min}} \cdot \frac{1}{N_Z} \tag{18}$$

This dependence is visualised in figure 1(b), from which it is easy to see that for a measurement of $A_{FB}$ the angular acceptance of the endcaps ($20° \lesssim \theta \lesssim 40°$ and $140° \lesssim \theta \lesssim 160°$) is approximately as important as that of the barrel region ($40° \lesssim \theta \lesssim 150°$). In other words: adding the information of the endcap-detectors is approximately equivalent to doubling the statistics.

## 4 The experimental detection efficiency

The symmetry properties of $f$ and $g$ can in principle be altered by asymmetries in the detection efficiency. Corrections due to such asymmetries can be assumed to be small, because of cancellations in the ratio of particles to anti-particles. The derivation of the moments started by only using ratios in which the detection efficiency cancels in every angular bin separately. As long as the detection efficiency does not depend on the 'charge' $C$, the answer will still be correct, although it could be that the error estimate is slightly biased by the angular dependence of the weights. If the angular dependence of the efficiency $E(\cos\theta)$ can be determined, this information can be used to make the corrections $f \to Ef$ and $g \to Eg$. Then eq. (15) is again valid. If the detection efficiency is very asymmetric in $\cos\theta$, it could be necessary to use the full equation (13), including the correction $\propto A^2$.

## 5 An experimental background

When the signal is not pure, the background can often be assumed to be symmetrical in $\cos\theta$. Often such backgrounds give an extra contribution to the symmetrical part of distribution (3):

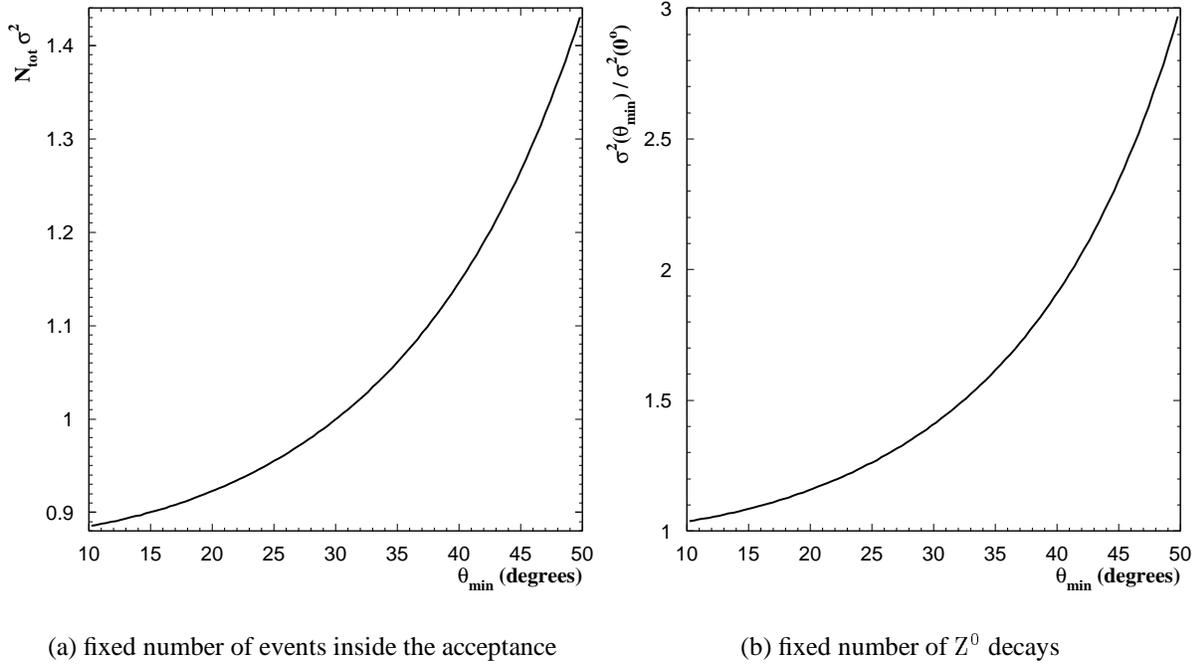

(a) fixed number of events inside the acceptance  (b) fixed number of $Z^0$ decays

**Figure 1:** *The dependence of the standard deviation $\sigma$ on the polar acceptance. Figure (a) visualises the coefficient of $1/N$ in equation (17). This gives the expected statistical error for a given number of detected events inside the angular acceptance. Figure (b) shows the angular dependence of $\sigma^2$ from equation (18), normalised to the case of full acceptance ($\theta_{\min} = 0$). This second plot shows the relative importance of different angular regions of a detector. For detectors at LEP, the borders between the 'barrel' and 'endcap' regions are near to $40°$ and $140°$.*

$f \to (1+D) = f$. We can call $D$ the 'dilution', while $R = 1/(1+D)$ is the 'purity': the fraction of the original signal in the total sample. After a renormalization of the distribution one can see that the detected asymmetry decreases as $A \to A/(1+D) = RA$. Let the measured asymmetry before correcting for the background be given by $A_0 \pm s/\sqrt{N}$. The accuracy of the corrected result will now depend on the number of events, the purity $R$ and on its uncertainty $\delta R$:

$$A = RA_0 \qquad \sigma_A^2 = \frac{s^2}{NR^2} + (A_0 \delta R)^2 \qquad (19)$$

This shows that it is more important to have a high and well known purity $R \approx 1$ than to have a high detection efficiency (large $N$).

Of course it is also possible that the background is not not proportional to $f(\cos\theta)$, or is not even a symmetrical function of $\cos\theta$. In this case the symmetry properties of $f$ (and possibly also $g$) can be modified. Such behaviour of the background can be taken into account by using the more complete relation (13). However, this is only possible after a precise determination of the background distribution.

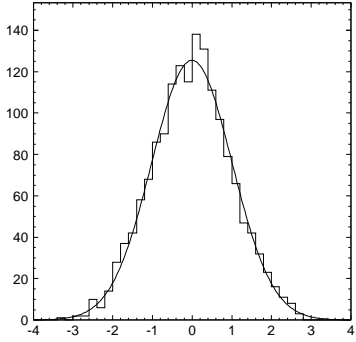

**Figure 2:** *The distribution of pulls $\delta A/\sigma$ for a measurement of $A$ using the method of moments on samples of $1600$ events. The curve shows the result of a fit to a gaussian distribution, with $\langle \delta A \rangle = -0.005 \pm 0.026$ and $\sigma = 1.012 \pm 0.020$.*

# 6  A Monte Carlo evaluation of the method

With a simple Monte Carlo program, the method of moments was compared to the unbinned likelihood method. The probability of the likelihood method was taken to be proportional to $N^+(C\cos\theta)$ as given by eq. (2). The distribution as generated in the Monte Carlo could be different from this, to simulate an angular dependence in the efficiency or the purity. For different numbers of Monte Carlo events one can determine the measured value of $A$ for both methods and the estimated standard deviations. Such a Monte Carlo measurement can be repeated many times. This makes it possible to determine the real standard deviations as well.

An internal check of both methods and of the program itself is to look at the distribution of the 'pulls': the difference between the input asymmetry and the measured values, divided by the estimated standard deviation. As it should be, the distribution of the pulls can be fitted very precisely by a gaussian centred at zero and with a width equal to $1$ to an accuracy of about $1\%$ (see e.g. figure 2).

It turns out that the method of moments is equivalent to the unbinned likelihood method: on average, the standard deviations are equal. Both methods are very robust against uncertainties in the detection efficiency. The difference between the two methods is less than a few percent of the error estimate for measurements using numbers of events ranging from $50$ to $256000$.

# 7  Conclusion

The method of moments developed here is equivalent to the unbinned likelihood method, but simpler to use and it gives useful analytical formulas for the expected statistical errors in certain simple but quite realistic cases. This is as good as one can get, since it is well known that the unbinned likelihood method is, nearly by definition, equivalent to the best possible method [1]. Both methods can be refined to include the effects of the efficiency and the background.